\documentclass[twocolumn,showpacs,aps,prl,superscriptaddress]{revtex4}
\usepackage{multirow}
\usepackage{graphicx}
\usepackage{dcolumn}
\usepackage{amsmath}
\usepackage{epsfig}
\input{ogbabarsym}

\newcommand{\SLACPubNumber} {15772}
\newcommand{\LANLNumber} {1312.6800}

\def\figurebox#1#2#3{%
    \def\arg{#3}%
    \ifx\arg\empty
    {\hfill\vbox{\hsize#2\hrule\hbox to #2{\vrule\hfill\vbox to #1{\hsize#2\vfill}\vrule}\hrule}\hfill}%
    \else
    {\hfill\epsfbox{#3}\hfill}%
    \fi}

\begin{document}

\preprint{SLAC-PUB-\SLACPubNumber} 

\begin{flushleft}
SLAC-PUB-\SLACPubNumber\\
arXiv:\LANLNumber\ [hep-ex]\\[10mm]
\end{flushleft}

\title{{\large\bf\boldmath
Search for the decay $\mydecII$
\unboldmath}}

%
\author{J.~P.~Lees}
\author{V.~Poireau}
\author{V.~Tisserand}
\affiliation{Laboratoire d'Annecy-le-Vieux de Physique des Particules (LAPP), Universit\'e de Savoie, CNRS/IN2P3,  F-74941 Annecy-Le-Vieux, France}
\author{E.~Grauges}
\affiliation{Universitat de Barcelona, Facultat de Fisica, Departament ECM, E-08028 Barcelona, Spain }
\author{A.~Palano$^{ab}$ }
\affiliation{INFN Sezione di Bari$^{a}$; Dipartimento di Fisica, Universit\`a di Bari$^{b}$, I-70126 Bari, Italy }
\author{G.~Eigen}
\author{B.~Stugu}
\affiliation{University of Bergen, Institute of Physics, N-5007 Bergen, Norway }
\author{D.~N.~Brown}
\author{L.~T.~Kerth}
\author{Yu.~G.~Kolomensky}
\author{M.~J.~Lee}
\author{G.~Lynch}
\affiliation{Lawrence Berkeley National Laboratory and University of California, Berkeley, California 94720, USA }
% ...
\collaboration{The \babar\ Collaboration}
\noaffiliation

\begin{abstract}
We report a search for the decay $\mydecII$. Using a data sample of 
$471 \times 10^6$ \BB pairs collected with the \babar~detector at the \pep2 ~storage 
ring at SLAC, we find no events and set an upper limit on the branching fraction
$\mybrII\times\frac{\BR(\Lcp\to\proton\Km\pip)}{0.050} <2.8\times10^{-6}$ at $90\,\%$ C.L.,
where we have normalized $\BR(\Lcp\to\proton\Km\pip)$ to the world average value.
\end{abstract}


\maketitle

\B mesons have approximately $7\,\%$ \cite{Beringer:1900zz} baryons among their
decay products. This is a substantial fraction justifying further investigations
that may allow better understanding of baryon production in \B decays and, more 
generally, quark fragmentation into baryons. The measurement of exclusive branching 
fractions of baryonic \B decays as well as systematic studies of the dynamics of the decay, 
i.e., the fraction of resonant subchannels, is a direct way for studying the mechanisms 
of hadronization into baryons.\\
\indent
We report herein a search for the decay $\mydecII$ \cite{CP} using a dataset of about 
$424 \invfb$ \cite{Lees:2013rw}, corresponding to $471 \times 10^6$ \BB pairs.
This decay is closely related to \thdec and $\mydecF$, which have a similar quark content 
and the (so far) largest measured branching fractions among the baryonic \B decays with a \Lcp~in the final state. 
The CLEO Collaboration has measured the branching fraction $\BR(\mydecF)=(23\pm7)\times10^{-4}$ \cite{Dytman:2002yd}.
\babar~reported a measurement of $\thbr=(12.3\pm3.3)\times10^{-4}$ as well as the branching ratios of resonant
subchannels with $\Sigma_{\rm c}(2455,2520)^{0,++}\to\Lcp\pi^{-,+}$ \cite{Lees:2013zm}.
The main differences between the decay presented here and the other two decay channels are 
the absence of possible resonant subchannels and the much smaller phase space (PS), e.g.,

\begin{equation}
\frac{\int d{\rm PS}(\mydecII)}{\int d{\rm PS}(\thdec)}\approx \frac{1}{1500} \,.
\end{equation}

Given the fact that the decay products of $\mydecII$ are limited to a small PS, a significant deviation from 
the phase space factor of $1/1500$ in the ratio of the branching fractions may occur if hadronization into 
$\Lcp\antiproton$ and/or $\proton\antiproton$ is enhanced due to their generally low invariant masses. 
This phenomenon is known as threshold enhancement and describes the dynamically enhanced decay rate at the 
baryon-antibaryon-mass threshold. It has been observed in baryonic \B decays with open charm final states 
\cite{Lees:2013zm,delAmoSanchez:2011gi,Aubert:2010zv,Aubert:2008ax}, charmless baryonic \B decays \cite{Wang:2003iz} 
and in the decay $\jpsi\to\g\proton\antiproton$ \cite{Bai:2003sw}. 
An example where the decay with the smaller PS is preferred is the ratio of $\BR(\Bub\to\Lcp\aLcp\Km)/\BR(\mjdecb)\approx3$ \cite{Beringer:1900zz}
with ${\int d{\rm PS}(\Bub\to\Lcp\aLcp\Km)}/{\int d{\rm PS}(\mjdecb)}\approx{1}/{70}$.
The influence of the weak interaction is expected to be similar here since $|V_{\c\s}|\approx|V_{\u\d}|$. 
General phenomenological approaches to describe the threshold enhancement consider, for example, gluonic and fragmentation 
mechanisms \cite{Rosner:2003bm} and pole models \cite{Suzuki:2006nn}.
In particular, an enhancement at the proton-antiproton mass threshold could be explained by the baryonium candidate $X(1835)$ 
\cite{Beringer:1900zz,Datta:2003iy,Ding:2005gh}. Other theorists propose the possibility of $S$ wave $\proton\antiproton$ 
final state interaction with isospin $I=1$ \cite{Sibirtsev:2004id} and contributions from one-pion-exchange interactions 
in $N\Nbar$ states with isospin $I=1$ and spin $S=0$ \cite{Zou:2003zn}. 

On the other hand, the decay \mydecII may be suppressed by the absence of resonant subchannels, which may play 
an important role for baryonic \B decays, e.g., the resonant part of $\thdec$ due to $\Sigma_{c}$ baryons is 
$\approx40\%$ \cite{Lees:2013zm}. The size of the branching fraction may allow us to balance the relevance of 
resonant subchannels against PS in baryonic \B decays.

This analysis is based on a data set that was collected with the \babar~detector at the 
\pep2 ~asymmetric-energy \epem storage ring, which was operated at a center-of-mass (CM) energy 
equal to the \FourS mass.
We use EvtGen \cite{Lange:2001uf} and \jetset74 \cite{Sjostrand:1993yb} for simulation of 
\babar~events, and GEANT4 \cite{Agostinelli:2002hh} for detector simulation. 
The sample of simulated decays \mydecII with $\Lcp\to\proton\Km\pip$, both uniformly
distributed in PS, is referred to as signal Monte Carlo (MC).

For the reconstruction of charged-particle tracks, the \babar~detector uses a tracking system
that consists of a five-layer double-sided silicon vertex tracker surrounding the beam pipe
followed by a 40-layer multiwire drift chamber with stereoangle configuration.
A superconducting solenoid generates an approximately uniform magnetic field of $1.5$ Tesla 
inside the tracking system which allows a precise measurement of the momentum of the tracks.
The selection of proton, kaon and pion candidates is based on measurements of the energy loss 
in the silicon vertex tracker and the drift chamber, and measurements of the Cerenkov radiation in the 
detector of internally reflected Cerenkov light \cite{Aubert:2002rg}. 
A detailed description of the \babar~detector can be found elsewhere \cite{Aubert:2001tu,TheBABAR:2013jta}. 

We reconstruct $\Lcp$ in the subchannel $\Lcp\to\proton\Km\pip$.
For the reconstruction of the \B candidate, we fit the entire $\mydecII$ decay tree simultaneously,
including geometric constraints to the \Bzb and \Lcp~decay vertices, and require the 
$\chi^2$ fit probability to exceed $0.1\,\%$.

Averaging over the momentum and polar angle of the particles that we use for our reconstruction
in the signal MC sample, the track finding efficiency is larger than $97\,\%$ \cite{Allmendinger:2012ch}.
The identification efficiency for protons and pions is about $99\,\%$ and for kaons about $95\,\%$
while the probability of a pion, kaon or proton to be misidentified is below $2\,\%$. In particular, the 
probability for a pion or kaon to be misidentified as a proton is negligible. Thus, we expect a low 
combinatoric background level due to the fact that three genuine protons originating from a common \B vertex, 
like for \mydecII, are rare in \babar~events.

To suppress background, we develop selection criteria for the \mydecII~ and \lcpdec~ candidates 
using correctly reconstructed decays in the signal MC sample.

\begin{figure}
\begin{center}
	\includegraphics[width=0.49\textwidth]{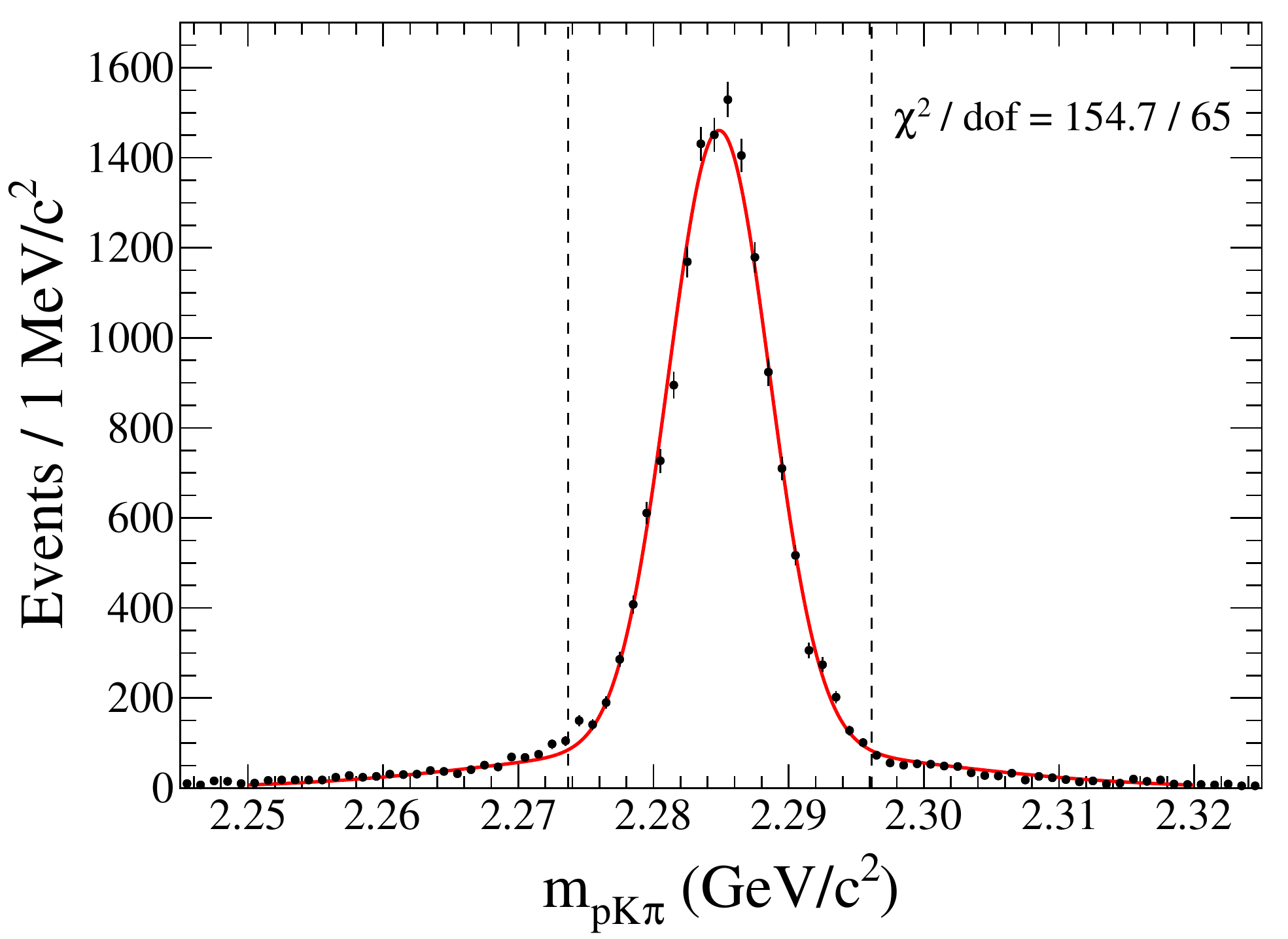}
	\caption{Fitted $m_{\proton\kaon\pi}$ distribution of correctly reconstructed \B events in signal MC. 
	The two dashed vertical lines enclose the $m_{\proton\kaon\pi}$ signal region corresponding to $\pm\,3\,\sigma$.}
	\label{fig:mLc}
\end{center}
\end{figure}

For $\proton\Km\pip$ combinations from \Lcp~decays, we observe a narrow and a broad signal 
component in the $m_{\proton\kaon\pi}$ invariant-mass distribution, in which the broad component results 
from badly reconstructed candidates. Thus, we fit the $m_{\proton\kaon\pi}$ invariant-mass distribution to a 
sum of two Gaussian functions with a common mean value (Fig.~\ref{fig:mLc}). We extract a 
standard deviation ($\sigma$) of $(3.74\pm0.04)\mevcc$ for the narrow component and $(15.4\pm0.4)\mevcc$ 
for the broad component. The uncertainty is purely statistical. The fraction of the narrow part 
is approximately $80\,\%$. The mean value ($\mu$) of $(2284.85\pm0.04)\mevcc$, that corresponds 
to our reconstructed mass, is in agreement with the generated \Lcp~mass used in the simulation.
To improve the signal-to-background ratio, we use only the signal region around the \Lcp~defined by
$\pm3\sigma$ of the narrow Gaussian function. For this selection, we achieve an efficiency of $89\,\%$.
We validate our method by reconstructing the \Lcp~decay inclusively in the \babar~data. For comparison we only 
select $\proton\kaon\pi$ combinations whose momentum is inside the momentum range of $\Lcp$ from correctly 
reconstructed $\mydecII$ decays in the signal MC sample.
We find that the widths and fractions of the fitted distribution to $m_{\proton\kaon\pi}$ from \Lcp~decays in the data 
sample and the signal MC sample are in agreement but the mean value is shifted by $0.5\mevcc$. 
Thus, the signal region in the $m_{\proton\kaon\pi}$ distribution in data is shifted correspondingly.

\begin{figure}
\begin{center}
	\includegraphics[width=0.49\textwidth]{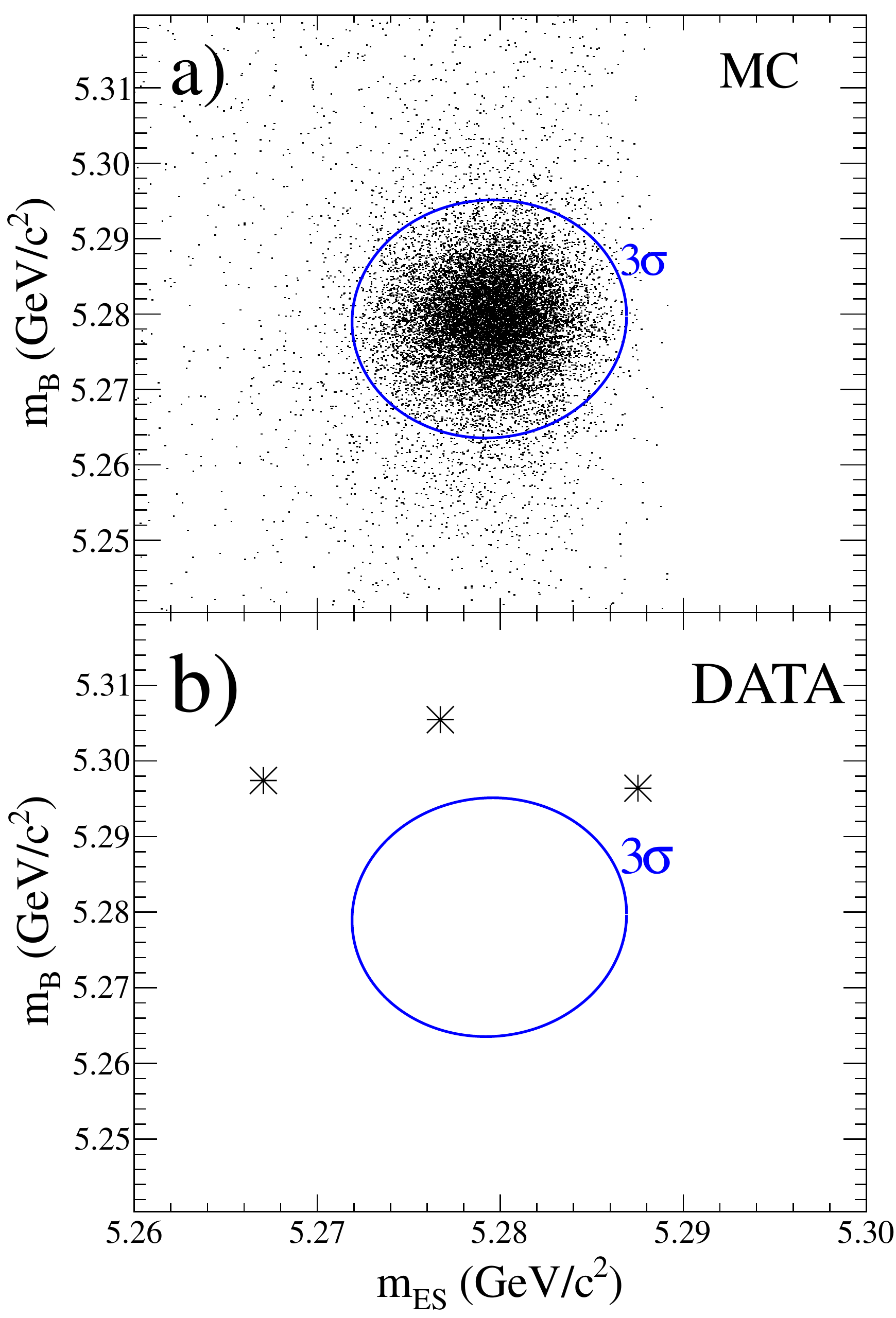}
	\caption{The \mes vs \mB distribution of selected events in (a) signal MC and (b) data.
	No signal candidates are observed within the signal region of the data sample.}
	\label{fig:mes-mb}
\end{center}
\end{figure}

The separation of signal from background in the \B candidate sample is obtained using
the invariant mass \mB and the energy-substituted mass 
$\mes = \sqrt{(s/2+\mathbf{p}_{\rm i} \cdot \mathbf{p}_{\rm \B}\cdot c^2)^2/E_{\rm i}^2-\left(\left|\mathbf{p}_{\rm \B}\right|\cdot c\right)^2}/c^2$, 
where $\sqrt{s}$ is the CM energy of the \epem pair. $(E_{\rm i}, \mathbf{p}_{\rm i})$ is the 
four-momentum vector of the \epem CM system and $\mathbf{p}_{\B}$ the \B-candidate momentum vector, 
both measured in the laboratory frame. For correctly reconstructed \B decays, \mB and \mes are centered
at the \B meson mass.
Figure~\ref{fig:mes-mb}(a) shows the \mes vs \mB distribution for all reconstructed \B candidates, 
including the selection criteria for $m_{\proton\kaon\pi}$. Both \mes and \mB peak at the nominal \Bzb 
meson mass and have a correlation coefficient of $2.6\%$.

We define a signal region for \Bzb decay candidates in the \mes-\mB plane that lies within a $3\sigma$ 
covariance ellipse around the nominal \Bzb mass [Fig.~\ref{fig:mes-mb}\,(a)]. Beside the correlation coefficient, 
the ellipse is defined by the mean value ($\mu$) and the standard deviation ($\sigma$) of both variables, 
whose determination is described in the following section. The prefix ``$3\sigma$''~refers to the fact that the 
length of the two half-axes of the ellipse is three times the $\sigma$ value of $\mes$ and $\mB$, respectively.

We fit a single Gaussian function to the \mes invariant-mass distribution yielding a mean of 
$\mu(\mes)=(5279.44\pm0.03)\mevcc$ and a standard deviation of $\sigma(\mes)=(2.62\pm0.02)\mevcc$.
As in the $m_{\proton\kaon\pi}$ case, the \mB invariant-mass distribution has both a narrow and broad component and we 
fit it to a sum of two Gaussian functions with a common mean. We obtain a mean of 
$\mu(\mB)=(5279.34\pm0.05)\mevcc$ consistent with the nominal $\Bzb$ mass. The narrow component contains
$80\,\%$ of the signal and has a standard deviation of $\sigma(\mB)=(5.26\pm0.07)\mevcc$ while that of 
the broad component is $\sigma(\mB)=(14.5\pm0.5)\mevcc$. The uncertainties again are purely statistical.
The selection in \mes and \mB, using the described signal region, has an efficiency of $82\,\%$.

To validate the viability of our selection in the \mes-\mB plane, we perform studies in the control 
channels $\B\to\Dbar^{(*)}\D^{(*)}\kaon$ \cite{PhysRevD.83.032004} and \thdec \cite{Lees:2013zm}.
For both decay channels, we find that the distributions of \mes vs \mB in the signal MC and in 
the data are in agreement, confirming that our MC is able to describe the data correctly.

Figure~\ref{fig:mes-mb}(b) shows the distribution of \mes vs \mB for \mydecII candidates in the data sample. 
Only three events remain after the selection in the vicinity of the signal region, and we do not 
observe any events inside the signal region.

We determine the selection efficiency from the number of reconstructed events in the signal MC sample
inside the signal region normalized to the total number of generated events; 
this yields an efficiency of $\varepsilon=(3.52\pm0.05)\,\%$.
We estimate the systematic uncertainty on the efficiency by repeating the analysis in the \mes vs \DeltaE plane, 
where $\DeltaE=E_{\rm\B}^*-\sqrt{s}/2$ is the deviation from the nominal energy of the \B candidate in the 
\epem CM system. As before, we define a $3\sigma$ signal region, where we find no \B candidates in the data sample, 
and determine the selection efficiency in the signal MC sample. The absolute difference in efficiencies is $0.02\,\%$. 
In addition, we account for the statistical uncertainty in the efficiency of $0.03\,\%$ resulting from the limited size 
of the signal MC sample. Furthermore, we estimate the uncertainty in the efficiency from tracking to be $0.03\,\%$.
Summing these values in quadrature, we determine a total uncertainty of $0.05\,\%$.
Other systematic uncertainties that influence the measurement of the branching fraction are due to uncertainties 
on the number of \B events and particle identification efficiency. We find these values to be negligible compared 
to the uncertainty of the reconstruction efficiency in the signal MC sample.

In Eq.~(\ref{eq:BF}), we define a modified branching fraction ($\BR_{\rm mod}$),

$$\BR_{\rm mod} = \mybrII \times \dfrac{\BR(\Lcp\to\proton\Km\pip)}{0.050}\,,$$

\begin{equation}
\label{eq:BF}
= \dfrac{N_{\rm observed}}{\varepsilon\cdot N_{\B}\cdot0.050}\,,
\end{equation}

which is the usual product branching fraction normalized to the world average value of 
$\BR(\Lcp\to\proton\Km\pip)=(0.050\pm0.013)$ \cite{Beringer:1900zz}. $N_{\rm observed}$ 
is the number of signal events and $N_{\B}=471\times10^{6}$ is the number of \Bzb mesons 
in the \babar~data set, assuming equal production of \BzBzb and \BpBm by \FourS decays.
The definition is equivalent to \mybrII but independent of the large external uncertainty 
on the branching fraction for \Lcp\to\proton\Km\pip.

In a Bayesian approach, we evaluate the probability density function (pdf) of $\BR_{\rm mod}$ 
given by $N_{\rm observed}$ and $\varepsilon$ by performing pseudoexperiments and 
determine an upper limit at $90\,\%$ C.L.
We vary the value of $N_{\rm observed}$ and $\varepsilon$ according to the following 
distributions:

\begin{equation}
\label{eq:Nobs}
P(x=N_{\rm observed}) = \left[\dfrac{x^{n}}{n!} \cdot e^{-x}\right]_{n=0} = e^{-x}\,,
\end{equation}

\begin{equation}
\label{eq:Eff}
P(\varepsilon) = \dfrac{1}{\sqrt{2\pi}\sigma(\varepsilon)} \cdot \exp\left[-\dfrac{1}{2}\left(\dfrac{\varepsilon-\mu(\varepsilon)}{\sigma(\varepsilon)}\right)^2\right].
\end{equation}

\begin{figure}[ht!]
\begin{center}
	\includegraphics[width=0.49\textwidth]{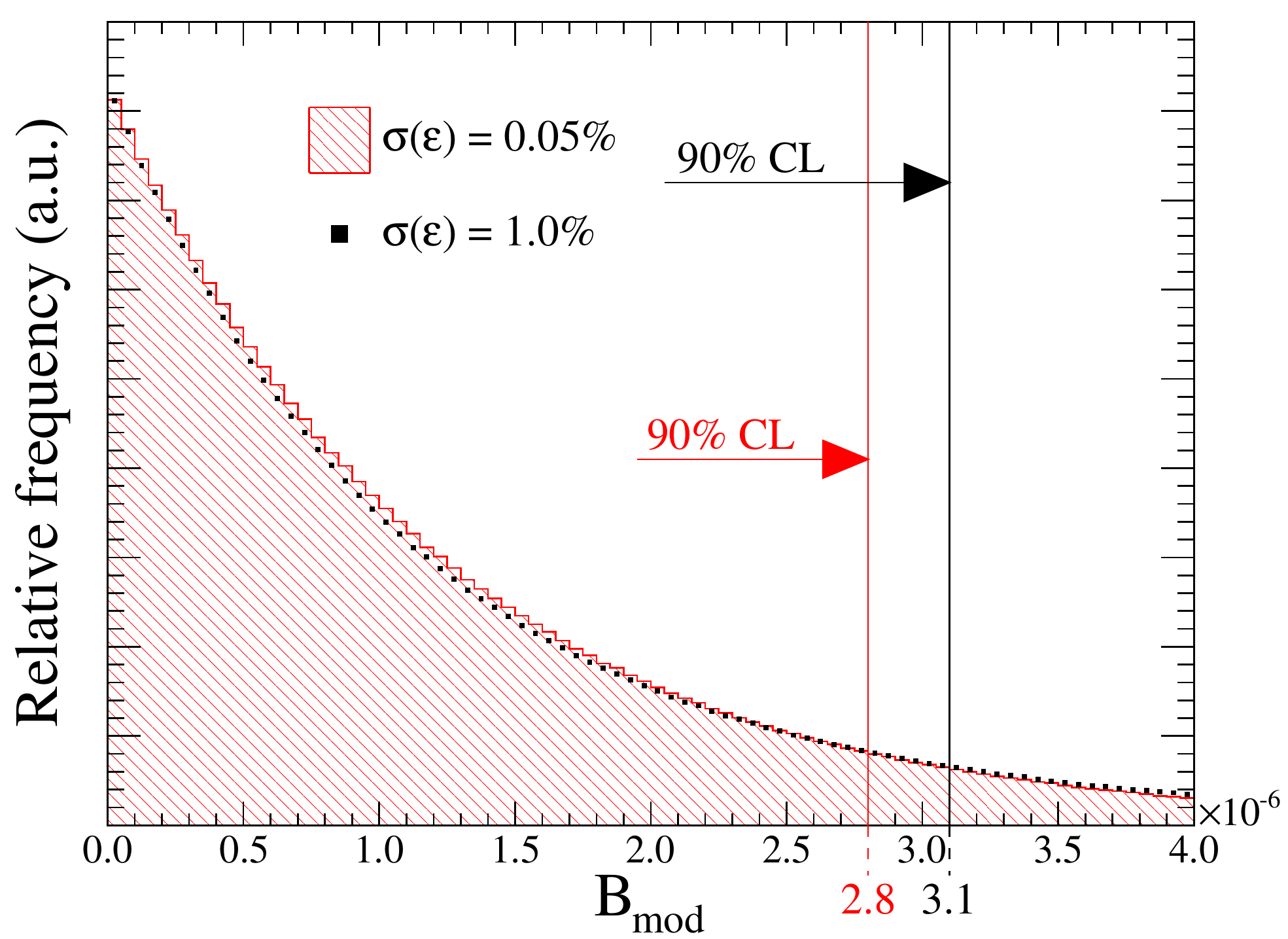}
	\caption{The distribution of the modified branching fraction $\BR_{\rm mod}$ 
	from pseudoexperiments using the calculated uncertainty of the reconstruction efficiency 
	$\sigma(\varepsilon)=0.05\,\%$ (hatched) and a $20$ times higher uncertainty $\sigma(\varepsilon)=1.0\,\%$ (squares).
	The two vertical lines indicate the upper limit of the $90\,\%$ C.L., respectively.}
	\label{fig:3}
\end{center}
\end{figure}

Equation~(\ref{eq:Nobs}) is a Poisson distribution that describes the pdf for finding no signal events ($n=0$) 
given by the true number of \mydecII decays $(x)$.
Equation~(\ref{eq:Eff}) represents a Gaussian distribution that models the pdf of the reconstruction efficiency.
We use the determined efficiency as the mean value $(\mu)$ and the uncertainty on the efficiency as the standard deviation $(\sigma)$.
Figure~\ref{fig:3} shows the distribution of $\BR_{\rm mod}$ for the given uncertainty of $\sigma(\varepsilon)=0.05\,\%$ and for a $20$ times higher 
uncertainty of $\sigma(\varepsilon)=1.0\,\%$ to assess the impact of systematic uncertainties on this quantity. We determine branching 
fraction upper limits at the $90\,\%$ confidence level of $BF<2.8\times10^{-6}$ for $\sigma(\varepsilon)=0.05\,\%$ and 
$BF<3.1\times10^{-6}$ for $\sigma(\varepsilon)=1.0\,\%$, respectively. 
The upper limit rises to $2.9\times10^{-6}$ only at $\sigma(\varepsilon)=0.55\,\%$.
This demonstrates that our result is robust against systematic uncertainties.

To summarize, we have searched for the decay \mydecII using a sample corresponding to an 
integrated luminosity of $424\invfb$ in \epem collisions at the \FourS resonance, 
collected with the \babar~detector. We find no events and derive the upper limit on 
the branching fraction,

\vspace{-1em}

$$\mybrII \times \dfrac{\BR(\Lcp\to\proton\Km\pip)}{0.050}$$

\vspace{-1.5em}

\begin{equation}
< 2.8 \times 10^{-6} \ \text{at }90\,\% \text{ C.L.,}
\end{equation}

\noindent
where we normalize the product branching fraction to the world average value of 
$\BR(\Lcp\to\proton\Km\pip) = 0.050$. 
We interpret the upper limit on $\BR(\mydecII)$ in comparison to the nonresonant branching fraction of $\BR(\thdec)$.
We use the result $\BR(\thdec)_{\rm non-\Sigma_c}=(7.9\pm2.1)\times10^{-4}=\left(0.64\pm0.17\right)\cdot\BR(\thdec)$ 
published in \cite{Lees:2013zm}. In addition, we take into account contributions from additional intermediate states 
including $\Deltabar^{--}$ and $\rho$ resonances that are not accounted for in the analysis, but that are visible in 
the invariant mass spectra of $\antiproton\pim$ and $\pip\pim$. In summary, we estimate that 
$\BR(\thdec)_{\rm non-res} \approx 0.5\cdot\BR(\thdec)$. Therefore, we calculate 

\begin{equation}
\label{eq:res} 
\frac{\BR(\mydecII)}{\BR(\thdec)_{\rm non-res}}\lesssim\frac{1}{220}\ .
\end{equation}

If we separate the dynamic and kinematic factors that contribute to the branching fraction according to \\
$\BR \sim \int \left|M\right|^2 \cdot d{\rm PS} = \langle \left|M\right|^{2}\rangle\times \int d{\rm PS}$, where 
$\langle \left|M\right|^{2}\rangle = \frac{\int \left|M\right|^2 d{\rm PS}}{\int d{\rm PS}}$ 
is the average quadratic matrix element of the decay, we can write 

\begin{equation}
\label{eq:concl}
\frac{\BR(\mydecII)}{\BR(\thdec)_{\rm non-res}}=r\times\frac{1}{1500}\ .
\end{equation}

In Eq.~(\ref{eq:concl}) we applied $\frac{\int d{\rm PS}\left(\mydecII\right)}{\int d{\rm PS}\left(\thdec\right)}=\frac{1}{1500}$
and introduced an effective scaling factor $r$ that quantifies the enhanced production rate of baryons due to dynamic effects.
Using the result from Eq.~(\ref{eq:res}) we obtain

$$r=\frac{\langle \left|M\left(\mydecII\right)\right|^{2}\rangle}{\langle \left|M\left(\thdec\right)\right|^{2}\rangle}\lesssim6.8 \ .$$

This is in tension with the quantities \\ $\BR(\Bub\to\Lcp\aLcp\Km)/\BR(\mjdecb)\approx3$ and \\ 
${\int d{\rm PS}(\Bub\to\Lcp\aLcp\Km)}/{\int d{\rm PS}(\mjdecb)}\approx\frac{1}{70}$,\\ which leads to a factor of $r=210$ 
without subtracting contributions from intermediate states in \mjdecb.
Under the used assumptions we conclude that a significantly enhanced decay rate of 
$\mydecII$~w.r.t.~$(\thdec)_{\rm non-res}$ due to dynamic effects that are related to the threshold enhancement 
does not exist.

We are grateful for the 
extraordinary contributions of our \pep2 \ colleagues in
achieving the excellent luminosity and machine conditions
that have made this work possible.
The success of this project also relies critically on the 
expertise and dedication of the computing organizations that 
support \babar.
The collaborating institutions wish to thank 
SLAC for its support and the kind hospitality extended to them. 
This work is supported by the
US Department of Energy
and National Science Foundation, the
Natural Sciences and Engineering Research Council (Canada),
the Commissariat \`a l'Energie Atomique and
Institut National de Physique Nucl\'eaire et de Physique des Particules
(France), the
Bundesministerium f\"ur Bildung und Forschung and
Deutsche Forschungsgemeinschaft
(Germany), the
Istituto Nazionale di Fisica Nucleare (Italy),
the Foundation for Fundamental Research on Matter (The Netherlands),
the Research Council of Norway, the
Ministry of Education and Science of the Russian Federation, 
Ministerio de Ciencia e Innovaci\'on (Spain), and the
Science and Technology Facilities Council (United Kingdom).
Individuals have received support from 
the Marie-Curie IEF program (European Union) and the A. P. Sloan Foundation (USA).


\begin{thebibliography}{99}
\bibitem{CP} 
 The use of charge conjugate decays is implied throughout this paper.
\end{thebibliography}
\end{document}